\documentclass{elsart}
\usepackage{graphicx}
\usepackage{amssymb}
\usepackage{amsmath}

\journal{arXiv}

\begin{document}
\begin{frontmatter}



\title{A constitutive model for volume-based descriptions of gas flows}

\author{S.\ Kokou Dadzie}
\ead{kokou.dadzie@strath.ac.uk} \and
\author{Jason M.\ Reese\corauthref{cor}}
\ead{jason.reese@strath.ac.uk}
\corauth[cor]{Corresponding author.}
\address{Department of Mechanical Engineering,
University of Strathclyde,\newline Glasgow G1 1XJ, UK}

\begin{abstract}
We derive the pressure tensor and the heat flux to accompany the new
macroscopic conservation equations that we developed in a
volume-based kinetic framework for gas flows \cite{vol-kokou}. This
kinetic description allows for expansion or compression of a fluid
element, and is characterized by a flux of volume that conventional
modelling does not account for. A new convective form of transport
arises from this new approach, which is due purely to macroscopic
expansion or compression. It supplements the conventional transport
processes, which are transport due purely to mass motion (classical
convective transport) and transport due purely to random motion
(diffusive transport). In addition, we show that the thermodynamic
parameters of the fluid (temperature and pressure) appear with new
non-equilibrium contributions that depend on density variations in
the gas. Our new model may be useful for describing gas flows that
display non-local-thermodynamic-equilibrium (rarefied gas flows),
flows with relatively large variations of macroscopic properties,
and/or highly compressible fluids and flows.
\end{abstract}

\end{frontmatter}

\section{Introduction}
The conventional fluid mechanics description is the set of three
Navier-Stokes equations, which can be retrieved from the kinetic
theory of rarefied gases using, for example, a perturbation method
(known as the Chapman-Enskog expansion) truncated at first order
\cite{cercignaniblanc,chapman}. The validity of the Navier-Stokes
set of equations is predicated on small deviations from
thermodynamic equilibrium, which can be interpreted physically as
relatively slow and low amplitude variations of macroscopic
variables within the flows. Describing flows beyond the broad range
of applicability of the Navier-Stokes model (such as hypersonic
flows, or micro or nano scale gas flows) remains an active area of
investigation
\cite{LockerbyReeseGallis2005b,reesegallislockerby2003,koga1954-Chem.Phy,struchtrupbook,myong2004}.

An intuitive way of constructing hydrodynamic models beyond the
range of applicability of Navier-Stokes is high-order development of
the Chapman-Enskog expansion. These are the Burnet and Super-Burnet
equations. However, these high-order expansions have various and
well-known physical inconsistencies, including frame-dependent
diffusive fluxes \cite{wood-JFM1983}, violation of the second law of
thermodynamics \cite{Comeaux-AIAA1995}, and other instabilities
\cite{bobylev-1982}. Various ways of overcoming these
inconsistencies have been proposed
\cite{jinslemrod2001,Agarwal-2001}.

In a previous paper \cite{vol-kokou}, we discussed the
representation of fluids and flows from both the gas kinetic and the
classical fluid mechanical viewpoints. We introduced a kinetic
representation of dilute gases based on volume variations. The
fundamental change introduced in our new model was that, while gases
are physically represented by particles undergoing random motions
and collisions, the free volume around each gaseous molecule should
be taken into account in their microscopic modeling, and that this
free microscopic volume around each particle is linked to
macroscopic variables such as the density of the fluid. We modified
Boltzmann's original kinetic equation, which is a simple balance of
the number of particles, to account for the microscopic free volume
around particles. We therefore arrived at a new set of macroscopic
conservation equations. In addition, two different macroscopic
velocities, leading to two different definitions of kinetic peculiar
velocities, were distinguished: the mass-velocity (which is the
velocity of the centre of mass of an element of fluid), and the
volume-velocity (which allows for expansion/compression in the
motion of a fluid element).

The aim of the present paper is to build on this new volume-based
description by developing first-order constitutive relations for the
pressure tensor and  heat flux to accompany it.

This paper is structured as follows:

Section \ref{sec-sum} summarises our new volume-based kinetic
approach.

In section \ref{sect-new-marco-convect}, an unconventional
convective form of transport arising from our new kinetic approach
is presented. This convective transport is due purely to macroscopic
expansion/compression, and supplements the two conventional
transport processes, namely a transport due purely to mass motion
(the classical convective transport) and a transport due to random
motion (without any bulk motions, and which is called diffusive
transport).

In section \ref{sec-cons} we derive new constitutive relations. This
derivation is mainly based on the identification of diffusive fluxes
as convective fluxes in our new framework. We argue that actual
diffusive fluxes are those due purely to particle random motions, so
they should be fluxes from which not only mass motion but also
macroscopic expansion/compression motion contributions should be
removed. A complete new set of hydrodynamic equations is then
proposed.

In section \ref{sect-last}, we analyse the new quantities
involved in our proposed hydrodynamic model. We show that the flux
of volume, which is normally a convective flux, behaves also like a
diffusive flux. Finally, the new thermodynamic parameters generated
by our model are compared with the thermodynamic parameters of the
classical description.

(Notation: In this paper, the square of the modulus of a vector $Z$
we denote by $Z^2$.)

\section{A volume-based kinetic description of gas flows \label{sec-sum}}

In the kinetic theory of simple monatomic gases, the gaseous
particles are only characterized (at any time $t$) by their
position, $X$, and velocity, $\xi$. In order to describe situations
sensitive to variations of macroscopic parameters, especially
density, we introduced the microscopic free volume, $v$, around each
gaseous particle into the set of microscopic parameters
characterizing the molecules \cite{vol-kokou}. A distribution
function $f(t, X,\xi, v )$ is then a probability number density of
particles which, at a given time $t$, are located in the vicinity of
position $X$, have their velocities in the vicinity of velocity
$\xi$, and have an assigned microscopic volume in the vicinity of
volume $v$.

As the dynamics of the gaseous particles is described through their point-masses,
the kinetic equation describing the evolution of $f(t, X, \xi, v
)$ is the following Boltzmann-like equation:
\begin{equation}
\label{eq.boltzmann.kok}
\frac{\partial f}{\partial t} + (\xi \cdot \nabla ) f +(F \cdot
\nabla_\xi ) f + W  \frac{\partial f}{\partial v} = \int \int (f^*
f_1^* - f f_1) \sigma \xi_r d_{\omega} d_{\xi_1} ,
\end{equation}
in which the term on the right-hand-side is the classical
Boltzmann collision integral; $f = f(t, X, \xi, v )$ and $f_1 = f(t,
X, \xi_1, v_1 )$ refer to post-collision particles, $f^* = f(t, X,
\xi^*, v^* )$ and $f_1^* = f(t, X, \xi_1^*, v_1^* )$ refer to
pre-collision particles, $\xi_r = \xi-\xi_1 $ is the particle
relative velocity, $\sigma$ the collision differential cross
section, $d_\omega$ an element of solid angle, and $\nabla_\xi$
denotes the formal operator $\nabla_\xi =\partial/\partial \xi_x
+\partial/\partial \xi_y +
\partial/\partial \xi_z $. On the left-hand-side
appears a new term involving $W$, which arises from the introduction
of the new variable $v$ into the distribution function. This term
represents the internal rate of change of volume occupied by the gas
\cite{vol-kokou}. In the following we will suppose no body force,
i.e.\ $F=0$ (although situations of non-vanishing body force can be easily
integrated into the description).

\subsection{The macroscopic parameters of the gas}
The macroscopic parameters according to our modified distribution function $f(t, X, \xi, v )$
are defined as follows:

\begin{itemize}
    \item
    The local number-density of the molecules is given by
\begin{equation}
n(t,X) =  \int_{-\infty}^{+\infty}  \int_0^{+\infty} f(t, X, \xi, v
)  d_v d_\xi \ .
\end{equation}
This number density refers to the fixed observer's reference frame with
coordinate elements $(X_1,X_2,X_3)$, in which any motion and
dynamics are described.

\item
    The local mean value, $\bar{Q}(t,X)$, of any property $Q$ can be defined by,
\begin{equation}
\label{meanvalu}
n(t,X) \bar{Q}(t,X)   =  \int_{-\infty}^{+\infty}  \int_0^{+\infty} Q  f(t, X, \xi, v )
d_v d_\xi \ .
\end{equation}
For example, the local mean microscopic volume, $\bar{v}(t,X)$, around each gaseous particle
is given by,
\begin{equation}
\label{meanvoldefi} n(t,X) \bar{v} (t,X) = \int_{-\infty}^{+\infty}
\int_0^{+\infty} v f(t, X, \xi, v )  d_v d_\xi \  .
\end{equation}

\item
From this mean value of the microscopic volume, a local mean value
of the mass-density in the vicinity of position $X$ is defined
through:
\begin{equation}
\label{massdensitydef}
\bar{\rho} (t,X)  = \frac{n(t,X) M }{n(t,X)\bar{v}(t,X)} =
\frac{M}{\bar{v}(t,X)}  \ ,
\end{equation}
where $M$ is the molecular mass.
\end{itemize}

It may be noted that in gas flows where the molecules can be assumed
always uniformly dispersed (locally), the mean value of the
microscopic volumes, $\bar{v}$, is simply given by  $d_X/(nd_X )=
1/n$, regardless of definition (\ref{meanvoldefi}). In this case the
definition of density, equation (\ref{massdensitydef}), becomes
$\bar{\rho} = Mn$. However, we are interested here in general
situations where these restrictions on the volume occupied by a
group of particles are relaxed; this is in order to get more
efficient descriptions of flows with density variations. While the
density $Mn$ can be regarded as the density of a gas without density
variations, the actual mass-density of the general gas (i.e.\ the
mass divided by the actual volume occupied by the gas) is given by
$\bar{\rho}$ in equation (\ref{massdensitydef}).

Two mean velocities can be defined. First, the local mean mass-velocity, $U_m (t,X)$, is given through
\begin{equation}
\label{vitessemass}
M n(t,X) U_m (t,X) = \int \int M  \xi f(t, X, \xi, v ) d_\xi d_v .
\end{equation}
A local mean volume-velocity, $U_v(t,X)$, can also be defined:
\begin{equation}
\label{vitessevolume}
\bar{v} (t,X) n(t,X)  U_v (t,X) = \int \int  v \xi f(t, X, \xi, v ) d_\xi d_v .
\end{equation}
While $U_m (t,X)$ is purely the velocity of the
centre of mass of a ``fluid particle'' (or an element of fluid) around
$X$, the velocity $U_v (t,X)$ takes into account any
expansion or compression velocity of this fluid particle.
This is because the weighting element in equation (\ref{vitessevolume}),
which is a volume, is a microscopically changing property in contrast to the mass
in equation (\ref{vitessemass}).

From these two local mean velocities, two peculiar velocities can therefore be introduced.
The usual mass-velocity definition of peculiar velocity is
\begin{equation}
\label{peculiarmass} C  = \xi - U_m \ .
\end{equation}
But another peculiar velocity may be given through the
volume-velocity, i.e.
\begin{equation}
\label{peculiarvol} C'  = \xi - U_v \ .
\end{equation}

\subsection{A new set of macroscopic conservation equations}

Multiplying  the modified kinetic equation (\ref{eq.boltzmann.kok})
by the four microscopic quantities $v$, $M$, $(M\xi)$, $(M\xi^2)$
and then integrating over $d_v$ and $d_\xi$, we obtained the
following set of macroscopic conservation equations:
\begin{description}
\item[Continuity]
\begin{equation}
\label{massnewmacro} \frac{D n }{ Dt} = -  n \nabla \cdot U_m  \ ,
\end{equation}
\item[Mass-density]
\begin{equation}
\label{densitynewmacro} \frac{D \bar{\rho} }{ D t}
=\frac{\bar{\rho}^2}{M }\left [ \frac{1}{n }\nabla \cdot
[\mathbf{J}_v] - W \right] ,
\end{equation}
\item[Momentum]
\begin{equation}
\label{momentumnewmacro} n \frac{D U_m }{ D t}  = -  \nabla \cdot
\mathbf{P} ,
\end{equation}
\item[Energy]
\begin{equation}
\label{energymacro}
 n \frac{D }{ D t} \left[\frac{1}{2} U_m^2 + e_{in}\right]
= - \nabla \cdot  \left[ \mathbf{P}\cdot U_m\right] - \nabla \cdot
\mathbf{q} \ .
\end{equation}
\end{description}
where the material derivative $D/Dt \equiv
\partial /\partial t + U_m \cdot~\nabla$.

In this set of equations $\mathbf{J}_v$, $\mathbf{P}$, $e_{in}$ and
$\mathbf{q}$ simply denote the following quantities:
\begin{equation}\label{Jv}
\mathbf{J}_v = \int \int v C f  d_v d_\xi \ ,
\end{equation}
\begin{equation}
\label{momentum-flux} \mathbf{P}_{ij}= \int \int C_i C_j f d_v d_\xi
\ ,
\end{equation}
\begin{equation}
\label{int-energy}
  n e_{in}= \frac{1}{2} \int \int  C^2 f d_\xi d_v \ ,
\end{equation}
\begin{equation}
\label{enrgy-flux}
  \mathbf{q} = \frac{1}{2} \int \int C^2 C f d_\xi d_v \ .
\end{equation}
It is important to note that these quantities, which are all written
with the peculiar velocity $C$ based on the mass-velocity $U_m$, are
simply quantities appearing through the rational mathematical
derivation of the macroscopic equations. They are not yet shown to
be the same as the pressure tensor, heat flux, or internal energy in
the fluid. In particular, the presence here of $C$ in equations
(\ref{momentum-flux}) to (\ref{enrgy-flux}) is due to the definition
within equation (\ref{vitessemass}) and to the Boltzmann integral
collision which vanishes when $(M\xi)$ and $(M\xi^2)$ are introduced
in the integral forms of equation (\ref{eq.boltzmann.kok}).

Our new set of macroscopic conservation equations contains four
equations instead of the usual three. Equation (\ref{massnewmacro})
simply ensures the conservation of mass. It involves a
number-density, $n$, which should be regarded as accounting only for
the number of particles, and the mass-velocity, $U_m$. On the other
hand, equation (\ref{densitynewmacro}) gives the evolution of the
actual density $\bar{\rho}$ as a thermodynamic parameter of the
fluid which varies with the volume occupied by the fluid.

\section{An unusual convective transport driven
by a flux of volume \label{sect-new-marco-convect}}

Using the definitions of the mean velocities and peculiar
velocities, equations (\ref{vitessemass}) to (\ref{peculiarvol}),
the following identities can be easily proved:
\begin{equation}
\label{prop-vol-velo1} \int \int   C' f d_\xi d_v  =  - n(U_v - U_m)
\ , \qquad
\int \int  v C' f d_\xi d_v  = 0  \ ,
\end{equation}
and
\begin{equation}
n \bar{v}  U_v = \int \int  v U_m  f d_\xi d_v   + \int \int v C f
d_\xi d_v  \ .
\end{equation}
So equation (\ref{Jv}) may be re-written:
\begin{equation}
\label{relationvitesmassvol}
\mathbf{J}_v  = n \bar{v}  \left( U_v -  U_m \right) \    .
\end{equation}
In equation (\ref{relationvitesmassvol}), $\mathbf{J}_v$ is
characterizing a macroscopic motion defined by ($U_v - U_m$), given
that $U_m$ and $U_v$ are both macroscopic velocities by definition.
The flux $\mathbf{J}_v$ therefore acts like a convective flux in
which the convected element is the volume transported at the
macroscopic velocity ($U_v - U_m$).

\begin{figure}
\begin{center}
\includegraphics{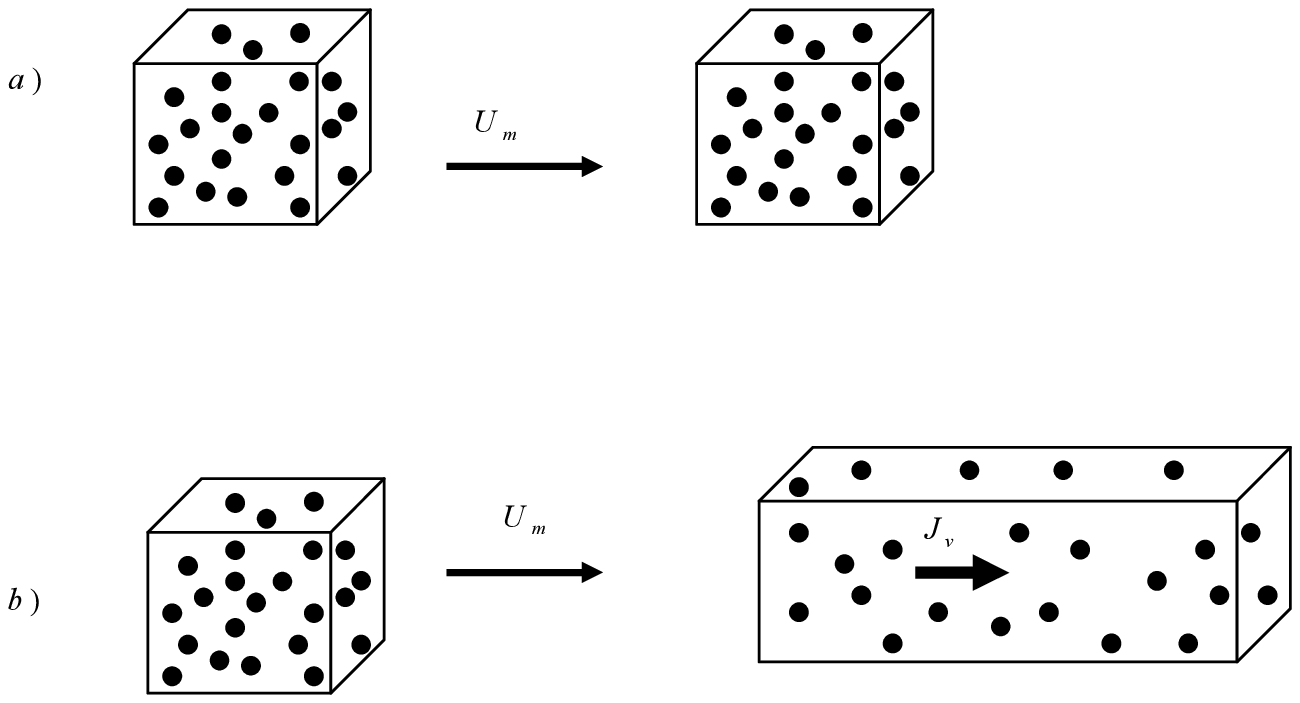}
\caption{Schematic of fluid element mass velocity, $U_m$, and volume
velocity, $U_v$; $a)$ the macroscopic motion of a fluid element in a
fluid without density variations, $b)$ macroscopic motion in a fluid
with density variations, $U_v  = U_m + 1/(n\bar{v})\mathbf{J}_v $.
Each of the volumes contain the same number of molecules.
\label{vitesse-constitu}} \vspace{1.5em}
\end{center}
\end{figure}

This unconventional macroscopic motion represented by $\mathbf{J}_v$
corresponds to a macroscopic change of volume of a ``fluid
particle'' in hydrodynamic descriptions. A fluid particle is not a
simple point-mass, but a volume element; this is illustrated in
Figure \ref{vitesse-constitu}.
If $\delta V$ represents the macroscopic change of volume of a fluid
element over time $\delta t$, then an approximate
expansion/compression speed is given by, $ \parallel
(n\bar{v})^{-1}\mathbf{J}_v\parallel  \approx   \delta
(V^{1/3})/\delta t \ , $ where $\parallel . \parallel $ denotes the
modulus of a vector. The velocity $\mathbf{J}_v$ is oriented in the
direction from high density to low density in the case of an
expansion, and vice versa for compression.

The macroscopic motion characterized by the flux $\mathbf{J}_v$
affects the classical interpretation of ``convective fluxes''. In
our volume-based description, the term ``convective flux'' should
include the macroscopic motion defined through the flux
$\mathbf{J}_v$, as well as the traditional centre of mass velocity
$U_m$.

\section{Constitutive relations in our volume-based description \label{sec-cons}}
In this section we identify the various fluxes to be associated with
the pressure tensor and heat flux, and the expression of internal
heat energy. Then we derive constitutive relations for these various
quantities, and for the flux $\mathbf{J}_v$.

\subsection{An expression for the flux of volume}

Let us regard the microscopic volume, $v$, as a property attached to
the molecules in a gas that has no macroscopic motion (or
macroscopic expansion or compression). Then equation (\ref{Jv}), as
it stands, represents a diffusive flux in which the microscopic
element undergoing diffusion is $v$. By a diffusive flux we
understand a flux due purely to the random motions of the particle,
without any form of bulk motion. In this case, following
phenomenological laws of diffusion \cite{Nelson}, we may expect the
diffusive volume flux to have the form
\begin{equation}\label{expresJ_v}
\mathbf{J}_v = - \kappa_v
 \  \nabla \bar{v} \ ,
\end{equation}
or, equivalently, by using equation (\ref{massdensitydef}),
\begin{equation}
\label{expresJ_v-rho}
 \mathbf{J}_v  = \frac{\kappa_v \bar{v }}{\bar{\rho}} \  \nabla \bar{\rho}  \ ,
\end{equation}
where we will call $\kappa_v$ the ``volume diffusion
coefficient''.

On the other hand,  we showed in the previous section that the flux
$\mathbf{J}_v$ physically characterizes the macroscopic motion
displayed in equation (\ref{relationvitesmassvol}). Any flux which
is given by this macroscopic motion should be taken as a convective
flux. So, the diffusive-like expression for $\mathbf{J}_v$ in
equation (\ref{expresJ_v}) may be interpreted as a ``fictitious''
diffusion of volume. However, we resolve this dual nature of
$\mathbf{J}_v$ by explicitly proving in section \ref{sect-last} of
this paper that, in fact, in addition to defining a new convective
transport, $\mathbf{J}_v$ corresponds to a particular diffusion of
mass. That it has both a convective and a diffusive nature justifies
equation (\ref{expresJ_v-rho}).

\subsection{The pressure tensor, heat flux and internal energy}

Classical kinetic theory attributes the pressure tensor and heat
flux to fluxes deriving from the peculiar velocity
\cite{cercignaniblanc,chapman}. It is also generally understood that
any macroscopic motion is defined only by the mass-motion, $U_m$. So
purely random components of the partitioning motion would then be
justifiably defined by the peculiar velocity $C$ through equation
(\ref{peculiarmass}). The concept of pure macroscopic
expansion/compression motions, as pointed out in section
\ref{sect-new-marco-convect}, does not exist.

In the conceptual framework of hydrodynamics (and kinetic theory), a
diffusive flux should be a purely internal flux of momentum or
energy in which contributions due to any macroscopic motions are
removed. This requirement is not satisfied by the three quantities
defined in equations (\ref{momentum-flux}) to (\ref{enrgy-flux}).
The macroscopic motion that arises from equation
(\ref{relationvitesmassvol}), and then the convective contributions
due to $\mathbf{J}_v$, are still inherent in fluxes expressed using
the peculiar velocity $C$.

However, all macroscopic motions --- including macroscopic expansion
and compression of a fluid element --- are removed from the peculiar
velocity $C'$ defined in equation (\ref{peculiarvol}) because this
equation can be re-written, using equation
(\ref{relationvitesmassvol}), as
\begin{equation}
\label{peculiarremove} C'  =  \xi - U_m
-\frac{1}{n\bar{v}}\mathbf{J}_v \ .
\end{equation}
Furthermore, equations (\ref{peculiarmass}), (\ref{peculiarvol}) and
(\ref{relationvitesmassvol}) yield
\begin{equation}
\label{diff-CC} C - C'  =  U_v - U_m =
\frac{1}{n\bar{v}}\mathbf{J}_v \ .
\end{equation}
We could therefore express the fluxes $\mathbf{q}$ and $\mathbf{P}$,
and the quantity $e_{in}$ from equations (\ref{momentum-flux}) to
(\ref{enrgy-flux}), which are simply quantities defined with the
peculiar velocity $C$, using the following decomposition:
\begin{equation}
\label{C-decompose} C = C' + ( C - C') = C' + (U_v - U_m) \ .
\end{equation}

Then $\mathbf{P}(t,X)$ from equation (\ref{momentum-flux}) with
equation (\ref{C-decompose}) becomes
\begin{equation}
\label{moment-fluxnew} \mathbf{P}_{ij}= \mathbf{P'}_{ij} - n
\frac{1}{(n\bar{v})^2}(\mathbf{J}_v \mathbf{J}_v)_{ij}  \ ,
\end{equation}
where $\mathbf{J}_v\mathbf{J}_v\equiv
(\mathbf{J}_v\mathbf{J}_v)_{ij}$ is the second order tensor obtained
by the product of the coordinate components of vector
$\mathbf{J}_v$, i.e.\ $(\mathbf{J}_v\mathbf{J}_v)_{ij}
=({\mathbf{J}_v})_i({\mathbf{J}_v})_j$, and $\mathbf{P'}$ is the
actual momentum flux tensor
\begin{equation}
\label{moment-fluxnew'} \mathbf{P'}_{ij}(t,X)= \int \int C'_i C'_j
f d_v d_\xi \ .
\end{equation}
A scalar pressure $p'$ can then be introduced through the sum of the
three diagonal terms of $\mathbf{P'}$:
\begin{equation}
\label{pressure-new}
   3p' = M \mathbf{P'}_{ii} \ .
\end{equation}

From equations (\ref{int-energy}) and (\ref{C-decompose}) $e_{in}$
becomes
\begin{equation}
\label{intenerg-eirew}
   e_{in}=  e'_{in} -  \frac{1}{2(n\bar{v})^2}\mathbf{J}_v^2 \ ,
\end{equation}
where the actual internal energy $e'_{in}$ is given as
\begin{equation}
\label{intenerg-new}
 n e'_{in} = \frac{1}{2} \int \int  C'^2 f d_\xi d_v  \ .
\end{equation}
A temperature $T'$ can also then be defined:
\begin{equation}
\label{temp-new}
  \frac{3}{2} kT' \ =  M e'_{in} \ ,
\end{equation}
and using equations (\ref{pressure-new}), (\ref{intenerg-new}) and
(\ref{temp-new}) we obtain
\begin{equation}
\label{pressure-pr}
 p'=nkT' \ .
\end{equation}

Using equation (\ref{C-decompose}), the flux
$\mathbf{q}$, defined in equation (\ref{enrgy-flux}), is
re-written
\begin{equation}
  \mathbf{q}(t,X) =    \frac{1}{2} \int \int \left[C' + (U_v - U_m)\right]^2 C' f d_\xi d_v
   + (U_v - U_m) \frac{1}{2} \int \int C^2  f d_\xi d_v \ .
\end{equation}
Then equations (\ref{prop-vol-velo1}), (\ref{diff-CC})
and (\ref{intenerg-eirew}) yield
\begin{equation}
\label{enrgy-rewflux1}
 \mathbf{q} =  \mathbf{q'}  +\frac{1}{n\bar{v}}\mathbf{P'} \cdot \mathbf{J}_v  -  \frac{1}{2\bar{v}}
 \frac{1}{(n\bar{v})^2}\mathbf{J}_v^2 \mathbf{J}_v
 + \frac{1}{\bar{v}}\left( e'_{in} -  \frac{1}{2(n\bar{v})^2}\mathbf{J}_v^2\right
  )\mathbf{J}_v\ ,
\end{equation}
which can be written
\begin{equation}
\label{enrgy-rewflux}
  \mathbf{q}=  \mathbf{q'}  +\frac{1}{n\bar{v}}\mathbf{P'} \cdot \mathbf{J}_v
   + \frac{1}{\bar{v}}\left( e'_{in} -  \frac{1}{(n\bar{v})^2}\mathbf{J}_v^2\right
   )\mathbf{J}_v\ ,
\end{equation}
where the actual energy flux vector, $\mathbf{q'} \equiv
\mathbf{q'}(t,X)$, is given by
\begin{equation}
  \mathbf{q'} =    \frac{1}{2} \int \int C'^2 C' f d_\xi d_v
  \ .
\end{equation}

The fluxes $\mathbf{q'}$ and $\mathbf{P'}$, and the quantity
$e'_{in}$, appear as new internal quantities with contributions due
to any bulk motions removed. The ``actual'' diffusive fluxes, by
which we mean those with any kind of convective fluxes removed
(including those described in section \ref{sect-new-marco-convect}),
are the new quantities $\mathbf{P'}$ for the momentum and
$\mathbf{q'}$ for the energy. In other words, diffusive fluxes in
our volume-based kinetic description are those expressed using the
peculiar velocity $C'$, in accordance with equation
(\ref{peculiarremove}), and not the fluxes appearing in equations
(\ref{momentum-flux}) to (\ref{enrgy-flux}) .

Equation (\ref{intenerg-eirew}) describes the total internal energy
of a fluid particle, $e_{in}$, as being composed of a ``heat
energy'', $e'_{in}$, and an elastic energy component driven by
$\mathbf{J}_v$. If we imagine a fluid particle (containing a certain
amount of mass) moving from a high density region to a low density
region in the fluid, this will cause a dilatation of the fluid
particle if the quantity of mass in the fluid particle is not
allowed to change. Particles may move out of or into a fluid
particle, but the amount of mass should remain the same, cf. Figure
\ref{vitesse-constitu}. According to the first law of
thermodynamics, the fluid particle heat energy must exclude any kind
of potential energy. In equation (\ref{intenerg-eirew}) the term in
$\mathbf{J}_v$ is a kind of potential energy because it is related
to the density gradient, and so represents the dilatation work done
by the fluid particle in the fluid by density variations. So the
actual heat energy should correspond to $e'_{in}$, not $e_{in}$.

The process we have just described does not exist within classical
fluid mechanics, where the total internal energy of the fluid
particle is associated only with the heat energy which is directly
related to the temperature. For a moving and expanding or
contracting fluid particle, the classical fluid mechanics definition
of internal energy does not agree with the first law of
thermodynamics conception of heat energy as it combines a potential
energy of the fluid particle with its actual heat energy. We show
below that our distinction of the heat energy of the fluid may
become sensitive in some extreme flow configurations (e.g. rarefied
gas flows under high temperature or density gradients).

\subsection{Constitutive model and the final set of hydrodynamic equations}

As the fluxes due  purely to random molecular motions in our kinetic
volume-based description correspond to $\mathbf{q'}$ and
$\mathbf{P'}$, it follows from phenomenological laws of diffusion
that these fluxes may be modelled at first order by linear
constitutive relations. These relations are the classical Newtonian
law for the stress tensor within the pressure tensor, $\mathbf{P'}$,
and Fourier's law for the heat flux, $\mathbf{q'}$. The
corresponding temperature and pressure in the fluid are then $T'$
and $p'$, respectively.

As a result, we obtain a set of hydrodynamic equations given by the
four macroscopic equations closed by equations
(\ref{moment-fluxnew}), (\ref{intenerg-eirew}) and
(\ref{enrgy-rewflux}), in which:
\begin{eqnarray}
\label{fluxes-new} M \mathbf{P'}_{ij} & = & p' \delta_{ij} -  \mu'
\left( \frac{\partial U_{v_i}}{\partial X_j} + \frac{\partial
U_{v_j}}{\partial X_i}\right) -\eta' \frac{\partial
U_{v_k}}{\partial X_k}\delta_{ij} \ ,
 \\ \nonumber
M \mathbf{q'}  & = & - \kappa'_h  \nabla T' \ ,  \\ \nonumber
 \mathbf{J}_v & = &  \frac{\kappa_v \bar{v }}{\bar{\rho}} \  \nabla \bar{\rho}  \ ,
\end{eqnarray}
with $\mu'$ a dynamic viscosity, $\kappa'_h$ a heat
conductivity, $\eta'$ a bulk viscosity, $\kappa_v$ a volume
diffusion coefficient, all to be determined in this volume-based framework.

The macroscopic motion which defines the momentum flux $\mathbf{P'}$
as a diffusive flux is $U_v$. It follows that the phenomenological
law of diffusion used to express $\mathbf{P'}$ should be applied
with a gradient taken over $U_v$, as we write in equations
(\ref{fluxes-new}). So momentum diffusion is not generated by the
gradient of the mass-velocity $U_m$ alone, but by the dilatation
motion $(U_v-U_m)$.

In the Appendix to this paper, we comment further on the expression
of the stress-tensor in equations (\ref{fluxes-new}) and compare it
to the common fluid mechanics stress tensor that involves only the
mass-velocity $U_m$.

The complete set of new hydrodynamic equations is rewritten below,
for convenience:
\begin{description}
\item[Continuity]
\begin{eqnarray}
\label{massnewhydro} \frac{D n }{ Dt} = -  n \nabla \cdot U_m \ ,
\end{eqnarray}
\item[Mass-density]
\begin{eqnarray}
\label{densitynewhydro} \frac{D \bar{\rho} }{ D t}
=\frac{\bar{\rho}^2}{M }\left [ \frac{1}{n }\nabla \cdot
[\mathbf{J}_v] - W \right] ,
\end{eqnarray}
\item[Momentum]
\begin{eqnarray}
\label{momentumnewhydro} n \frac{D U_m }{ D t}  = -  \nabla \cdot
\left(\mathbf{P'} -  \frac{1}{n\bar{v}^2}\mathbf{J}_v \mathbf{J}_v
\right) ,
\end{eqnarray}
\item[Energy]
\begin{eqnarray}
\label{energyhydro}
 n \frac{D }{ D t} \left[ \frac{1}{2} U_m^2 + e'_{in}  \right.
&& \left. - \frac{1}{2(n\bar{v})^2}\mathbf{J}_v^2 \right] =
\\ \nonumber && - \nabla \cdot  \left[ \left(\mathbf{P'} -
\frac{1}{n\bar{v}^2}\mathbf{J}_v \mathbf{J}_v \right)\cdot
U_m\right]
\\ \nonumber
 && - \nabla \cdot \left[\mathbf{q'}
+\frac{1}{n\bar{v}}\mathbf{P'} \cdot \mathbf{J}_v
   + \frac{1}{\bar{v}}\left( e'_{in} -  \frac{1}{(n\bar{v})^2}\mathbf{J}_v^2\right
   )\mathbf{J}_v\ \right] \ .
\end{eqnarray}
\end{description}
These equations are solved for five unknown fluid macroscopic
parameters: the number-density, $n$, the mass-density (which
involves the actual volume occupied by the gaseous particles),
$\bar{\rho}$, the mass-velocity, $U_m$, the pressure, $p'$, and the
internal energy, $e'_{in}$ (or, equivalently, the temperature,
$T'$). The equation set is then closed by the constitutive model
equations (\ref{fluxes-new}), and $M e'_{in}= (3/2) kT'$ , $p'=nkT'$
and $U_v = U_m + (n\bar{v})^{-1}\mathbf{J}_v$.

It is important to note that in the momentum equation
(\ref{momentumnewhydro}), the macroscopic velocity on the
left-hand-side is the mass-velocity, $U_m$, while on the
right-hand-side, according to equations (\ref{fluxes-new}), appears
the volume velocity, $U_v$. The modifications introduced by our new
hydrodynamic description, when compared to the usual Navier-Stokes
set of equations, recall the recent phenomenological modifications
of Brenner \cite{Brenner.PhysicaA.revs.2005}. That is to say,
replacement of the mass velocity by the volume velocity in (only)
the stress tensor, and the contribution to the heat flux of the work
done by any variation of volume.

In addition to Brenner's original modifications, second order terms
due to the volume flux appear in both the momentum and the energy
equations. Another difference between our model and Brenner's is our
distinction between the real mass-density of the fluid, which we
denote $\bar{\rho}$, and the mass-density of an equivalent
incompressible fluid, which is $Mn$ in our notation.

In contrast to classical kinetic theory, equation
(\ref{pressure-pr}) is no longer an equation of state, as $n$ does
not encompass the real volume of an element of fluid. This equation
will become a perfect gas equation of state only in the equilibrium
situation, where $Mn$ becomes $\bar{\rho}$, i.e.\ if no density
variation occurs. Variations of the density with the fluid
macroscopic parameters are described through equation
(\ref{densitynewhydro}), which is therefore functioning like an
equation of state.

An expression suggested for $W$ in our previous paper
\cite{vol-kokou}, leads to the following re-writing of the density
equation (\ref{densitynewhydro}):
\begin{equation}
\label{density-final}
\frac{1}{\bar{\rho}}\frac{D \bar{\rho}}{ D t} = \frac{\bar{\rho}}{Mn
}\nabla \cdot [\mathbf{J}_v] - \alpha \left(  \frac{D T'}{D t}  +
\frac{1}{n \bar{v}}\mathbf{J}_v  \cdot \nabla T'\right) + \chi\left(
\frac{D p'}{Dt}+ \frac{1}{n \bar{v}}\mathbf{J}_v  \cdot \nabla
p'\right) \ ,
\end{equation}
in which we may take $\alpha \approx 1/ T'$ and $\chi \approx 1/
p'$. It can be seen that when the flux $\mathbf{J}_v$ vanishes,
e.g.\ in the case of a (local) thermodynamic equilibrium, equation
(\ref{density-final}) is of the form of the classical perfect gas
equation of state. However, strictly speaking, the type of equation
of state which should be displayed by the density equation for a
vanishing flux of volume depends on how the rate of change of
volume, $W$, is described in the kinetic description
\cite{vol-kokou}.

In equations (\ref{massnewhydro}), (\ref{momentumnewhydro}) and
(\ref{energyhydro}), the presence of the number density $n$ (or the
apparently constant density $Mn$) on the left-hand-side, is founded
by a ``mass-based integration'' during the derivation of the
conservation equations \cite{vol-kokou}. This is interpreted as
following a fixed amount of mass when writing conservations of mass,
momentum and energy. This is consistent with the material derivative
appearing in all the left-hand-sides of these equations. The
evolution of the domain occupied by the fixed amount of mass is
contained in an equation of volume --- that is our equation of
mass-density that functions like an equation of state.

Finally, it can be easily seen that our new hydrodynamic model
becomes the Navier-Stokes set of equations when there is no
mass-density variation, i.e. when $\mathbf{J}_v =0$.

\section{Analysis of the new hydrodynamic parameters   \label{sect-last}}

\subsection{The flux of volume as a diffusion of the mass-density \label{massdiffusion}}

Here we show that $\mathbf{J}_v$ refers also to a particular mass
diffusion process that justifies the expression for it which
we hypothesised in equation (\ref{expresJ_v-rho}).

In our new volume-based kinetic description, all diffusive fluxes
have any bulk motion removed, including the expansion/compression
motion defined by $(U_v -U_m)$. These diffusive fluxes are therefore
associated with the peculiar velocity $C'$ given in equation
(\ref{peculiarremove}). Let us consider the following diffusive
flux,
\begin{equation}
\label{JmF} \mathbf{J}_{\bar{\rho}} = \int \int \bar{\rho} C' f d_v
d_\xi \ = \frac{1}{\bar{v}} \int \int M C' f  d_v d_\xi  \ .
\end{equation}
According to the definition of diffusive fluxes, the flux
$\mathbf{J}_{\bar{\rho}}$ corresponds to a diffusive flux of mass;
strictly speaking, the element undergoing diffusion in this equation
is the mass-density $\bar{\rho}$.

Using the definitions of the mean velocities, equations
(\ref{vitessemass}) and (\ref{vitessevolume}), with the definition
of the peculiar velocity $C'$ in equation (\ref{peculiarvol}), it can be
shown that
\begin{equation}
\label{prop-vol-veloF}  \mathbf{J}_{\bar{\rho}}  =  -
\bar{\rho}n(U_v - U_m) \ .
\end{equation}
On the other hand, from equation (\ref{diff-CC}) we have
\begin{equation}
\label{diff-CCF} U_v - U_m = \frac{1}{n\bar{v}}\mathbf{J}_v \ .
\end{equation}
Combining equations (\ref{prop-vol-veloF})  and (\ref{diff-CCF}), we
obtain
\begin{equation}
\label{JvF-Jm}
 \frac{1}{\bar{\rho}}\mathbf{J}_{\bar{\rho}} = -
 \ \frac{1}{\bar{v}} \mathbf{J}_v \ .
\end{equation}
Equation (\ref{JvF-Jm}) is a symmetrical relation between the flux
of volume, $\mathbf{J}_v$, and the diffusive flux
$\mathbf{J}_{\bar{\rho}}$.

As $\mathbf{J}_{\bar{\rho}}$ is a diffusive flux, according to
equation (\ref{JmF}), we can apply the phenomenological laws of diffusion:
\begin{equation}
\label{expresJ_m-pheno} \mathbf{J}_{\bar{\rho}} = - \kappa_m \nabla
\bar{\rho} \ ,
\end{equation}
where $\kappa_m$ is the mass diffusion coefficient (strictly speaking,
the mass-density diffusion coefficient).
From equation (\ref{JvF-Jm}) we then obtain,
\begin{equation}
   \ \mathbf{J}_v = \frac{\kappa_m\bar{v}}
{\bar{\rho}} \nabla \bar{\rho} \  ,
\end{equation}
which is the same as the expression for $\mathbf{J}_v$ written in
equation (\ref{expresJ_v-rho}), but with the volume diffusion
coefficient equal to the mass diffusion coefficient.

In equation (\ref{JmF}) or (\ref{expresJ_m-pheno}), we see that the
property of the fluid concerned in this diffusion is strictly the
mass-density. In other words, this diffusion process acts to smooth
out density inhomogeneities in a flow. This diffusion process should
not be confused with self-diffusion, which is normally considered as
a diffusion of matter through itself (in equilibrium).
Self-diffusion occurs in homogeneous matter, with zero mass-density
gradients, where one can imagine certain marked particles moving
through similar but unmarked particles \cite{chapman}.

Considering the mass-density as a ``concentration of mass within the
volume'', i.e. a mass divided by the volume of fluid, this diffusion
process could be describes as ``diffusion of mass through the volume
of fluid''. Or, from the symmetrical relation (\ref{JvF-Jm}),
``diffusion of the fluid volume through the mass''. If we consider a
gas at rest in which a density gradient is maintained, then this
diffusion process may be regarded as a constant flux of volume
imposed in the domain, or an apparent movement of the mass. On the
other hand, if imposed density gradient is suddenly relaxed, then
this diffusion corresponds to actual movement of mass from the high
mass-density regions to low mass-density regions.

Finally, regarding an element of fluid, any diffusive flux is
measured in a reference frame that moves not only with the mass
velocity but also undergoes expansion or compression with the
element of fluid. In this global reference frame, the expansion or
compression motion itself always looks like a diffusion of the
mass-density. This is consistent with the expansion or compression
associated with the flux $\mathbf{J}_v$ in earlier sections of this
paper. The coefficients $\kappa_m$ and $\kappa_v$, which have the
dimension of $(\mathrm{length})^2/\mathrm{time}$, resemble expansion
or compression coefficients, rather than purely diffusion
coefficients.

\subsection{Non-equilibrium thermodynamic parameters}

We now show that the scalars $T'$ and $p'$ defined in our
volume-based description correspond to the gas thermodynamic
parameters in non-equilibrium conditions. They coincide with the
conventional scalars, $T$ and $p$, only in equilibrium
conditions.

We introduce temporarily a scalar, $T$, which is related to the
total internal energy, $e_{in}$, in a similar way to equation
(\ref{temp-new}), and a scalar $p$ related to the sum of the three
diagonal terms of the total momentum flux tensor $\mathbf{P}$, in a
similar way to equation (\ref{pressure-new}), i.e.,
\begin{equation}
  \frac{3}{2} kT  =  M e_{in} \ ,
\end{equation}
and
\begin{equation}
   3p = M \mathbf{P}_{ii} \ .
\end{equation}
The quantities $p$ and $T$ would be the pressure and temperature if
the pressure tensor and heat flux were, as in classical kinetic
theory, connected to the fluxes defined with peculiar velocity $C$,
disregarding the convective transport involved by the flux
$\mathbf{J}_v$.

It can be shown in a similar way to that outlined previously, that
$p$ satisfies
\begin{equation}
\label{pres-show}
p=nkT \ .
\end{equation}
In addition, we obtain from equation (\ref{moment-fluxnew})
\begin{equation}
   3p = M \mathbf{P'}_{ii}-
Mn \frac{1}{(n\bar{v})^2}(\mathbf{J}_v \mathbf{J}_v)_{ii} \ ,
\end{equation}
which leads to
\begin{equation}
\label{pressuretot-new}
   p'  = p +  \frac{\bar{\rho}}{3n\bar{v}}\mathbf{J}_v^2 \ ,
\end{equation}
and so, from equations (\ref{pressure-pr}) and (\ref{pres-show}),
\begin{equation}
\label{temperaturetot-new}
   T'  = T +  \frac{\bar{\rho}}{3kn^2\bar{v}}\mathbf{J}_v^2 \ .
\end{equation}

It appears, therefore, that the pressure and temperature defined
within our new framework are different from the conventional
equilibrium scalars through equations (\ref{pressuretot-new}) and
(\ref{temperaturetot-new}). The difference may be attributed to the
contribution of mass-density variations to the pressure and the
temperature within the fluid, which in turn may be regarded as
non-equilibrium contributions to the pressure and temperature. The
term ``non-equilibrium'' associated with $T'$ and $p'$ refers to the
explicit contribution of an additional flux due to density
variations in their respective expressions, when compared with the
expressions for $T$ and $p$.

An expansion of gas during which the gas is allowed neither to
exchange heat with its surrounding nor to do external work is called
``free expansion'', and is an irreversible process. As the gas has
to expand into its allowed volume, work is done internally. Such an
internal process within a gas subject to density variations is
predicted with our new internal energy equation
(\ref{intenerg-new}): while the internal energy $e_i'$ is constant
(i.e.\ while the non-equilibrium temperature $T'$ is a constant),
internal work can be done through $\mathbf{J}_v$. This work done
within the gas contributes to changes in the thermal agitation of
the particles, i.e.\ changes in $T$. Moreover, in equations
(\ref{pressuretot-new}) and (\ref{temperaturetot-new}) the work term
due to $\mathbf{J}_v$ is similar to the classical thermodynamic work
[pressure $\times$ volume], but expressed in terms of density
variations, and the work done is transferred to the gaseous particle
agitation instead of into an external work. So, the new
contributions of $\mathbf{J}_v$ to temperature and pressure resemble
a ``free expansion'' heating in which the volume variation is given
by the local density gradient.

If the contribution of $\mathbf{J}_v$ is disregarded and the thermal
agitation temperature of the particles is taken as the thermodynamic
parameter and then related directly to the internal energy of the
particle, as in equation (\ref{temperatureclassic}) --- in other
words, if we take the classical fluid mechanics viewpoint --- then
constant internal energy implies constant temperature of the gas:
the free expansion effect should not exist. However, temperature
variations due to free expansion are well known and experimentally
demonstrated.

A stationary gas under uniform mechanical (not necessarily free)
expansion or contraction has $U_m=0$, so $\mathbf{J}_v =
n\bar{v}U_v$. Then equations (\ref{pressuretot-new}) and
(\ref{temperaturetot-new}) yield
\begin{equation}
\label{dilatationpressure}
   p' = p  + \frac{Mn}{3}U_v^2 \ ,
\end{equation}
and
\begin{equation}
\label{dilatationtemperature}
   T' = T  + \frac{M}{3k}U_v^2 \ .
\end{equation}
This means the total pressure of the gas during this expansion or
compression, when compared with the pressure within the same gas
under equilibrium, should be augmented by a contribution depending
on the square of the expansion speed. This additional pressure is a
type of dynamic pressure (i.e.\ the pressure given by $\bar{\rho}
U_m^2/2$), where the velocity involved is an expansion velocity
instead of the mass-velocity. Likewise, the actual temperature of
the fluid is not simply given by the heat energy corresponding to
its internal energy but is supplemented by the work done during
expansion or contraction.

It is the case that the differences appearing in equations
(\ref{pressuretot-new}) and (\ref{temperaturetot-new}) between the
standard and the new definitions of the thermodynamics properties,
will be negligible for most ordinary flows.

Two examples will help illustrate the new non-equilibrium
thermodynamic parameters. Let us first consider a mechanical
compression (or an expansion against a piston) of argon gas. The
molecular mass of argon is $M=6.63\times 10^{-26}$ kg and the
Boltzmann constant is $k=1.38\times 10^{-23}$ J/K. At a piston
compression speed equivalent to $U_v=50$ m/s, the difference in
temperature due to the work term in equation
(\ref{dilatationtemperature}) is $4.0$ K.

Let us now consider a shock wave in a rarefied gas flow, within
which a rapid variation of temperature (at constant pressure) exists
over a distance of the order of a molecular mean free path,
$\lambda_m$, which is of the same order of magnitude as the
characteristic length defined by the density gradient. In this case,
in equation (\ref{temperaturetot-new}) the expansion/compression
speed may be approximated by $\parallel
(n\bar{v})^{-1}\mathbf{J}_v\parallel  \approx \lambda_m/\tau_v \ $
 where $\tau_v$ is
the characteristic time of diffusion of volume (or mass-density).
Accordingly, the temperature difference in equation
(\ref{temperaturetot-new}) is
\begin{equation}
   \Delta T  = T'-T \equiv   \frac{M}{3k} \left(\frac{\lambda_m}{\tau_v} \right)^2\ .
\end{equation}
If we compare the diffusion time $\tau_v$ to the diffusion time for
energy, then the ratio $\lambda_m/\tau_v$ is close to the thermal
speed and we have $(\lambda_m/\tau_v)^2 \approx 3kT/M$.
Consequently, $\Delta T  \approx T$, i.e.\ the same order of
magnitude as the thermal speed temperature (the temperature commonly
identified with the flow). This difference is similar to some of the
temperature differences observed in high speed rarefied gas flows
during space vehicle re-entry, for which the classical theory of
surface temperature jump is known to have serious shortcomings
\cite{gupta-JTHT1996}.

\subsection{Interpretation of the new convective transport}

The two elementary transport processes admitted so far within fluids
are: the diffusion attributed to purely random motion of the
mass-particles, and transport due to the mass-motion or flow. The
additional transport process due to macroscopic expansion or
compression that we posit in this paper could be understood as
follows.

To an observer in a fixed reference frame, a fluid could be observed
to transport energy, mass, or momentum from one point to another by
simply undergoing expansion or compression. Evidently this cannot be
classified as a pure diffusion process, because the whole fluid
should be at rest. It cannot also be clearly identified with
classical convective transport due to the mass-motion, which is
physically the flowing of the fluid. This unclassified transport
process involves the evolution of the domain of the fluid --- it is
due to the change in volume of the fluid. It is a transport process
with characteristics of both elementary diffusion and convective
transport, but distinctively different from each; it is similar to
what is described in reference \cite{Brenner-Phys.vol2005} as a
``volume transport''.

\section{Conclusions}

In our previous paper \cite{vol-kokou} we reconsidered the
representation of fluids and introduced a new volume-based kinetic
approach for gases. The set of macroscopic conservation equations
derived from this approach comprises four equations rather than the
usual three. An evolution equation purely of the mass-density is
added to the set of three conservation equations. We therefore
distinguished between conservation of mass, which becomes a
conservation of number of particles, and the variation of the
mass-density as a thermodynamic parameter to account for any
variation in the volume occupied by an element of fluid. We also
distinguished between compressibility arising from fluid
accelerations and compressibility arising simply from the physical
properties of fluids.

In this present article we have proposed constitutive relations for
the pressure tensor and the heat flux to accompany our volume-based
macroscopic conservation equations. An unconventional convective
form of transport leads us to re-classify the diffusive and
convective fluxes. This extra convective mode of transport can be
viewed as a local macroscopic expansion or compression motion of the
fluid characterized by a flux of volume that classical modelling
does not account for.

As a result, we have obtained a hydrodynamic model in which the
contributions due to a flux of volume appear in the momentum and
energy equations. Furthermore, the thermodynamic parameters of the
fluid (temperature and pressure) appear with  non-equilibrium
contributions depending on density variations.

We propose that our model would show most departure from
conventional fluid mechanics in micro flow cases, where
compressibility effects are significant in spite of the low Mach
number. It could also be tested on flows which feature large density
and temperature variations, such as shock waves. Future work should
also include:
\begin{itemize}
\item investigating the boundary conditions to accompany our new set of
equations;
\item analysis of higher-order heat flux and pressure tensor derivations within our
new kinetic approach --- some classical expressions for these (such
as the Burnett equations) suffer from reference frame dependence,
but in our model the volume occupied by the molecules is treated
separately from the fixed reference volume element. This could
mitigate some of these problems and would, at least, yield
interesting new Burnett-like expressions.
\end{itemize}

\section*{Acknowledgements}
The authors would like to thank Howard Brenner of MIT (USA), Gilbert
M\'eolans of the Universit\'e de Provence (France), and Chris
Greenshields and Colin McInnes of Strathclyde University (UK) for
useful discussions. This work is funded in the UK by the Engineering
and Physical Sciences Research Council under grant EP/D007488/1, and
through a Philip Leverhulme Prize for JMR from the Leverhulme Trust.
JMR would also like to thank the Master and Fellows (in particular,
Prof Bill Fitzgerald) of Christ's College, Cambridge, for their
support and hospitality during a sabbatical year when this work was
completed.

\bibliography{constitu-relation}
\bibliographystyle{elsart-num}

\appendix

\section{The classical view of pressure tensor and heat flux}

In classical kinetic theory, macroscopic expansion or compression is
not involved in the flow description. It is understood that any
macroscopic motion is defined only by the mass-motion, $U_m$, and so
purely random motion would be defined by the peculiar velocity $C$
in equation (\ref{peculiarmass}). In this conceptual frame the
pressure tensor is given by $\mathbf{P}$ and the heat flux by
$\mathbf{q}$ \cite{cercignaniblanc,chapman}. Then, in accordance
with diffusion theory, these quantities are modelled by the
following standard constitutive relations:
\begin{eqnarray}
\label{fluxes} M \mathbf{P}_{ij} & = & p \delta_{ij} -  \mu \left(
\frac{\partial U_{m_i}}{\partial X_j} + \frac{\partial
U_{m_j}}{\partial X_i}\right) -\eta \frac{\partial U_{m_k}}{\partial
X_k}\delta_{ij} \ ,
 \\ \nonumber
M \mathbf{q}  & = & - \kappa_h  \nabla T  \ ,  \\ \nonumber
\end{eqnarray}
with $\mu$ the dynamic viscosity, $\kappa_h$ the heat conductivity,
$\eta$ the bulk viscosity \cite{chapman}. The stress tensor in
equations (\ref{fluxes}) is evidently written with the velocity
$U_m$, as the macroscopic motion associated with the peculiar
velocity $C$ is the mass velocity.

In fact, the stress tensor appearing in equations (\ref{fluxes})
does not account for all types of deformations within the fluid.
This is because $U_m$ misses any expansion/compression of a fluid
element. Strictly speaking, this stress tensor accounts only for the
``deformation of the mass-velocity''; it does not account for all
the deformations of a fluid element due purely to the changes in the
fluid physical properties, e.g. deformation due to heating and due
to density changes.

It has long been recognized that the deformation involved in
equations (\ref{fluxes}) takes place at constant temperature and
infers negligible effects from the elastic nature of fluids
\cite{malvern}. It cannot, therefore, efficiently describe flows in
which thermo-elastic effects are not negligible. Recently, Brenner
used experimental results of thermophoretic motion to suggest
replacing the velocity in the expression of the stress tensor by
another velocity that includes these missing deformations
\cite{Brenner.PhysicaA.revs.2005}. A similar remark can be made for
the heat flux expressed in equations (\ref{fluxes}) which also
appears to miss effects due to the elastic aspect of fluids.

The parts missing from the current fluid description appear to arise
from the treatment of elements such as a ``fluid particle'' and a
``control volume of fluid'' in the derivation of the fluid mechanics
set of equations. A fluid particle should not be treated as a simple
mass-point in which only the mass and the mass-velocity are
investigated: its volume, and changes in volume, should be taken
into account in the description.

Unlike the classical description, the new stress tensor obtained in
our volume-based kinetic approach encompasses the elastic nature of
the gaseous fluid, and allows for thermo-elastic deformations
in the fluid.

\end{document}